\documentclass[%
 reprint,
 amsmath,amssymb,
 aps,
]{revtex4-1}

\usepackage{graphicx}
\usepackage{dcolumn}
\usepackage{bm}
\usepackage[colorlinks,linkcolor=blue,anchorcolor=blue,citecolor=blue,urlcolor=blue]{hyperref}

\begin{document}

\title{Multi-phonon interactions between nitrogen-vacancy centers and nanomechanical resonators}

\author{Xing-Liang Dong}

\author{Peng-Bo Li}
\email{lipengbo@mail.xjtu.edu.cn}
\affiliation{Shaanxi Province Key Laboratory of Quantum Information and Quantum Optoelectronic Devices,
Department of Applied Physics, Xi'an Jiaotong University, Xi'an 710049, China}


\begin{abstract}
We investigate the multi-phonon interactions in a hybrid system composed of a nitrogen-vacancy
center and a mechanical resonator.
We show that, through appropriate sideband engineering analogy to the Mollow or Lamb-Dicke dynamics,
the enhanced nonlinear interactions can dominate the coupled system. As an example,
we show the preparation of nonclassical states of the mechanical motion
and explore the quantum correlation of $n$-phonon with the engineered dissipation,
based on the large multi-phonon coupling strength attainable in this sideband structure.
This work takes full advantage of the structure and coherence features of defect centers in diamond,
and may be useful for quantum information processing.
\end{abstract}

\maketitle


\section{introduction}

Quantum acoustics has attracted increasing attentions in recent years
\cite{Poot2012Mechanical,RevModPhys.86.1391}. This new field studies the quantum feature of mechanical degrees of freedom, and
their interactions with other systems, which not only furthers the fundamental physics, but also has a wide  application in quantum information processing. Plenty of acoustics devices, including mechanical resonators, surface acoustic wave devices, and acoustic waveguides, have been proposed and explored
\cite{cleland1996fabrication,Knobel2003Nanometre,Manenti2017Circuit,Satzinger2018Quantum,
Gustafsson207,PhysRevLett.120.213603,BandPeng,PhysRevApplied.10.064037}.
The available interactions with a range of quantum systems, such as superconducting qubits
\cite{LaHaye2009Nanomechanical,Chu199},
quantum dots
\cite{PhysRevLett.105.037401,Yeo2013Strain},
and solid-state defects
\cite{Doherty2013The,Lee2017Topical,Treutlein2014,
PhysRevLett.112.036405,PhysRevLett.118.223603},
open the possibility of coherent manipulation and control of mechanical motions.

On the other hand, several works based on different schemes  have succeeded in cooling the mechanical motion to its quantum ground state and making the thermal phonon population close to zero
\cite{PhysRevLett.92.075507,OConnell2010Quantum,Teufel2011Sideband,Chan2011Laser,
rabl2009strong,PhysRevB.82.165320,PhysRevB.88.064105},
allowing the preparation of more general quantum states through coherent or dissipative dynamics. So far, the stable entangled and squeezed states of mechanical motion are extensively investigated
\cite{PhysRevA.89.014302,PhysRevA.94.053807,Ockeloen-Korppi2018Stabilized,PhysRevA.79.063819,PhysRevA.91.013834}, and the generation of Schrodinger cat states of  motion is analyzed with trapped atoms
\cite{monroe1996schrodinger}.
However, the creation and coherent control of quantum states in nanomechanical systems is still a challenge
\cite{OConnell2010Quantum,chu2018creation}.

Solid-state defect color centers are one of the most appealing interfaces to mechanical oscillators due to their long coherence times
and perfect compatibility with other solid-state setups. In particular, single nitrogen-vacancy (NV) centers have an electronic spin ground state and possess the long coherence time even at room temperature
\cite{Childress281,Hanson2008Coherent,Balasubramanian2009Ultralong,Bar2013Solid}.
The interaction between NV centers in diamond and mechanical  modes has been actively discussed both in theory and experiment
\cite{rabl2009strong,PhysRevB.88.064105,Rabl2010A,PhysRevLett.117.015502,PhysRevA.88.033614,
Arcizet2011A,Kolkowitz1603,PhysRevX.6.041060,PhysRevLett.116.143602,PhysRevX.8.041027,PhysRevLett.121.123604,
PhysRevApplied.4.044003,PhysRevApplied.10.024011,PhysRevA.98.052346,PhysRevA.96.062333,PhysRevB.96.245418,PhysRevA.97.062318,
PhysRevA.99.013804,Cai:18}.
The realization of magnetic spin-mechanical couplings or excited-state mediated strain couplings permits the further study of more complicated dynamics process
\cite{PhysRevX.6.041060,rabl2009strong}.
Based on that, the generation of special nonclassical states of motion, for example the multi-phonon states, is very
useful and appealing.

In this work, we propose  to realize   multi-phonon interactions between  a strongly driven NV center and a mechanical resonator.
Through appropriate sideband engineering analogy
to the Mollow or Lamb-Dicke dynamics, the enhanced nonlinear interactions can dominate the
coupled hybrid system.
We study in detail the Mollow sideband
\cite{mollow1969br,PhysRevA.44.7717}
and Lamb-Dicke sideband
\cite{RevModPhys.75.281} transitions in this setup
\cite{Munoz2014Emitters,munoz2018filtering,N-phononQian}.
We reveal the multi-phonon interactions by getting an effective Hamiltonian. Different from the traditional method for creating $n$-phonon Fock states via  swap operations that are repeated $n$ times \cite{chu2018creation},
the sideband transitions we revealed  allow  to create $n$-phonon Fock states  through  just one swap operation,  due to the $n$-phonon interaction nature.   On the other hand, compared to the sideband analogy to the trapped ion schemes, we show that the Mollow sideband transition in this setup can get a much larger coupling strength, which offers much wider and more attractive applications.

When considering the multi-phonon interaction in the Mollow regime, we show that the system in the second sideband plus a nonlinearity can evolve into the motional cat states and realize the creation and manipulation of the Fock states directly by the coherent dynamical process. Moreover, we explore the $n$-phonon quantum correlation of the mechanical motion with a generalized correlation function. More applications and fundamental explorations can be carried out in this hybrid system with the attainable large coupling strengths.

\section{The setup}

\begin{figure}
\includegraphics[width=0.4\textwidth,trim=0 0 0 100,clip]{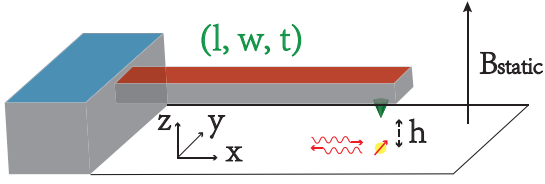}
\caption{\label{fig1}(Color online) Schematic of the setup.
A sharp magnetic tip is attached at the end of a silicon nanomechanical cantilever
of dimensions (l,w,t).
A single NV center is positioned below the tip at a distance of $h$,
and is immersed in a static magnetic field.
Additionally, microwave fields are added to drive the defect center.}
\end{figure}

We consider a hybrid system as shown in Figure~1,
where a nanomechanical cantilever couples to a single NV center with
an external magnetic tip  attached to the cantilever.
We place single NV centers at the distance of $h$ below the resonator.
We consider the case where the mechanical resonator oscillates  along the $z$-axis
which is aligned the NV symmetry axis.
In addition, a homogeneous magnetic field is applied to produce the splitting.
Furthermore, additional microwave (mw) fields are used to drive the spin.

NV centers are point defects in diamond with a substitutional nitrogen atom and an adjacent vacancy,
whose electronic ground state is a $S=1$ spin triplet state denoted as $|m_{s}=0,\pm 1\rangle$
\cite{Doherty2013The,Lee2017Topical}.
The existence of nonaveraged spin-spin interactions
causes a zero-field splitting $D=2\pi\times2.87$ GHz  with different values of $|m_{s}|$.
Moreover, the spin surrounded by a magnetic field
can get an interaction described by
$g_{s}\mu_{b}\hat{\vec{S}}\!\cdot\!\vec{B}_\text{tot}$,
with $g_{s}$ the Landé factor of NV center, $\mu_{b}$ the Bohr magneton
and $\hat{\vec{S}}$ the spin operator.
The cantilever is at the nanoscale and only the fundamental bending mode is considered.
Upon quantization, the free Hamiltonian of this mode is of $\hat{H}_{r}=\hbar\omega_{r}\hat{a}^\dag\hat{a}$.
Thus we can write the total Hamiltonian as $(\hbar =1)$
\begin{eqnarray}\label{ME1}
\hat{H}_\text{tot}=D{\hat{S}}_{z}^{2}+\omega_{r}\hat{a}^\dag\hat{a}+g_{s}\mu_{b}\hat{\vec{S}}\!\cdot\!\vec{B}_\text{tot}.
\end{eqnarray}

The total magnetic field consists of three parts:
the first from the mechanical oscillation of the resonator,
the second from the static field,
and the third from the classical mw fields, i.e.,
$\vec{B}_\text{tot}=\vec{B}_\text{tip}+\vec{B}_\text{static}+\vec{B}_\text{dr}$
with $\vec{B}_\text{static}=B_{0}\vec{e}_{z}$,
$\vec{B}_\text{dr}=B_{x}(t)\vec{e}_{x}+B_{z}(t)\vec{e}_{z}$.
In this setup, we expand the magnetic field $\vec{B}_\text{tip}$ up to first order.
The magnetic field induced by the vibrating tip has the form
$\vec{B}_\text{tip}=(B_{\text{tip}}(h)+G_{m}\hat{\text{z}})\vec{e}_{z}$,
where $G_{m}=\partial_{z}B_\text{tip}$ is the magnetic field gradient and
$\hat{\text{z}}=a_{0}(\hat{a}+\hat{a}^\dag)$
is the position operator,
with $a_{0}=\sqrt{\hbar/2m_{\text{eff}}\omega_{r}}$ the zero-point fluctuation amplitude,
$\hat{a}$ the annihilation operator,
and $\omega_{r}\sim2\pi\times5$ MHz the oscillating frequency of the resonator.
Then their interaction is expressed as $H_{\text{int}}=\lambda(\hat{a}+\hat{a}^\dag)\hat{S}_{z}$
with $\lambda=g_{s}\mu_{b}G_{m}a_{0}$.
In this work, we consider both the Mollow regime and the Lamb-Dicke regime with
different driving fields resulting in the same multi-phonon effective Hamiltonian.

\section{Multi-phonon interactions in the resolved sideband scheme}

\begin{figure}
\includegraphics[scale=0.2]{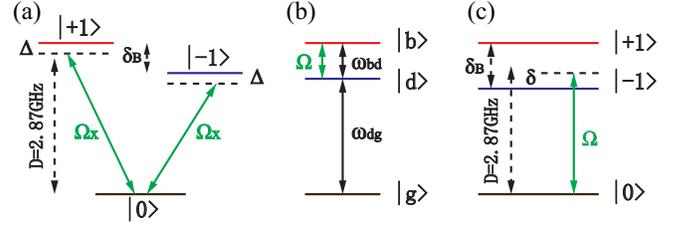}
\caption{\label{fig2a}(Color online) Level diagrams with respect to the Mollow and Lamb-Dicke sideband engineering schemes.
(a) The Level diagram of the NV center driven by two microwave fields.
The two driving fields have the same detunings and Rabi frequencies.
(b) Dressed spin states in the large detuning limit corresponding to (a).
The dressed states $|d\rangle$ and $|b\rangle$ are resonantly driven by microwave fields.
(c) The Level diagram of the NV center driven by a  weak microwave field.}
\end{figure}

\subsection{The Mollow sideband}

One of the most appealing sideband theory
primarily proposed by Mollow is known as Mollow sideband
\cite{mollow1969br,PhysRevA.5.2217}.
In that work, the resonance fluorescence spectrum of the strong driven two-level atom
reveals the high order interaction between
the strong driven two-level system (TLS) and the quantized fields.
Recently, the strong driven two-level atom in single mode optical cavity is well studied,
where the multi-photon transition in the Mollow sideband is displayed
\cite{Munoz2014Emitters,munoz2018filtering}.
In this section, with the  system of NV centers coupling to mechanical resonators,
we demonstrate the multi-phonon interaction with the sideband engineering analogy to the Mollow dynamics.

We apply three microwave fields in this scheme, $\vec{B}_\text{dr}=B_{x}(t)\vec{e}_{x}+B_{z}(t)\vec{e}_{z}$,
where $B_{x}(t)=B_{x0}^{(1)}\cos{\omega_{x}^{(1)}t}+B_{x0}^{(2)}\cos{\omega_{x}^{(2)}t}$,
and $B_{z}(t)=B_{z}\cos{\omega_{L}t}$.
With proper static magnetic fields, the total Hamiltonian can be rewritten as
\begin{eqnarray}\label{ME2}
\hat{H}_\text{tot}^{\text{M}}&=&D{\hat{S}}_{z}^{2}+\frac{\delta_{B}}{2}\hat{S}_{z}
+\omega_{r}\hat{a}^\dag \hat{a}\notag\\
&+&\lambda(\hat{a}+\hat{a}^\dag)\hat{S}_{z}+2\Omega\cos\omega_{L}t\hat{S}_{z}\notag\\
&+&2\sqrt{2}(\Omega_{x}^{(1)}\cos\omega_{x}^{(1)}t+\Omega_{x}^{(2)}\cos\omega_{x}^{(2)}t)\hat{S}_{x},
\end{eqnarray}
Here, $\delta_{B}=2g_{s}\mu_{b}(B_{0}+B_{\text{tip}}(h))$ is the splitting between the states
$|1\rangle$ and $|-1\rangle$,
$2\Omega=g_{s}\mu_{b}B_{z}$,
$2\sqrt{2}\Omega_{x}^{(1)}=g_{s}\mu_{b}B_{x0}^{(1)}$ and $2\sqrt{2}\Omega_{x}^{(2)}=g_{s}\mu_{b}B_{x0}^{(2)}$
are the driving amplitude.

Here, the two driving fields along the $x$ direction are used to dress the spin
with the same detunings $\Delta$ and Rabi frequencies $2\Omega_{x}$
($\Omega_{x}=\Omega_{x}^{(1)}=\Omega_{x}^{(2)}$), which is shown in Figure 2(a).
To show this effect, when $\Omega_{x}\gg\{\Omega,\lambda\}$,
we define the spin Hamiltonian as
\begin{eqnarray}\label{ME3}
\hat{H}_\text{spin}&=&D{\hat{S}}_{z}^{2}+\frac{\delta_{B}}{2}\hat{S}_{z}\notag\\
&+&2\sqrt{2}\Omega_{x}(\cos\omega_{x}^{(1)}t+\cos\omega_{x}^{(2)}t)\hat{S}_{x}.
\end{eqnarray}
Applying the unitary transformation
$U=\exp(i\omega_{x}^{(1)}t|1\rangle\langle 1|+i\omega_{x}^{(2)}t|-1\rangle\langle -1|)$
to Eq.~(\ref{ME3}),
we can acquire the Hamiltonian under the rotating wave approximation
\begin{eqnarray}\label{ME4}
\hat{H}_\text{spin}\simeq\sum_{j=\pm1}\Delta|j\rangle\langle j|
+\Omega_{x}(|0\rangle\langle j|+|j\rangle\langle 0|),
\end{eqnarray}
where $\Delta=D+\frac{\delta_{B}}{2}-\omega_{x}^{(1)}=D-\frac{\delta_{B}}{2}-\omega_{x}^{(2)}$
denotes the detuning.
Then, one of the dressed states is the dark state $|d\rangle=(|1\rangle-|-1\rangle)/\sqrt{2}$,
and in terms of the bright state $|b\rangle=(|1\rangle+|-1\rangle)/\sqrt{2}$ and the ground state $|0\rangle$,
we can obtain the other two dressed states
\begin{eqnarray}\label{ME5}
|g\rangle&=&\cos(\theta)|0\rangle-\sin(\theta)|b\rangle\\
|e\rangle&=&\sin(\theta)|0\rangle+\cos(\theta)|b\rangle\label{ME6},
\end{eqnarray}
with the auxiliary angle $\tan(2\theta)=2\sqrt{2}\Omega_{x}/\Delta$.
The corresponding eigenfrequencies are $\omega_{d}=\Delta$
and $\omega_{e/g}=(\Delta\pm\sqrt{\Delta^2+8\Omega_{x}^2}$)/2.
As a result, we can get the total Hamiltonian in the new basis as
\begin{eqnarray}\label{ME7}
\hat{H}_{\text{tot}}^{\text{M}}&\simeq&\omega_{r}\hat{a}^\dag \hat{a}+\omega_{eg}|e\rangle\langle e|
+\omega_{dg}|d\rangle\langle d|\notag\\
&+&[\lambda(\hat{a}+\hat{a}^\dag)+2\Omega\cos\omega_{L}t]
\times[-\sin(\theta)|g\rangle\langle d|\notag\\
&+&\cos(\theta)|d\rangle\langle e|+\text{H.c.}],
\end{eqnarray}
with $\omega_{eg}=\omega_{e}-\omega_{g}$ and $\omega_{dg}=\omega_{d}-\omega_{g}$.

In the large detuning limit ($\Delta\gg\Omega_{x}$), the dressed states are approximate to the bright state
$|b\rangle=(|1\rangle+|-1\rangle)/\sqrt{2}$, dark state
$|d\rangle=(|1\rangle-|-1\rangle)/\sqrt{2}$ and ground state $|g\rangle=|0\rangle$,
while a significant difference between $\omega_{bd}\approx2\Omega_{x}^2/\Delta$ and $\omega_{dg}\approx\Delta+2\Omega_{x}^2/\Delta$ can be observed [Fig.~2(b)].
Meanwhile, the value of $\cos(\theta)$ can approach one.
We take the states $\{|b\rangle, |d\rangle\}$ as the effective TLS
with $\hat\sigma=|d\rangle\langle b|$, and $\omega_{\sigma}=\omega_{bd}$.
When $\omega_{\sigma}\approx\omega_{r}$ and under the resonant driving ($\omega_{L}\approx\omega_{\sigma}$),
the system Hamiltonian can be reduced to the form as
\begin{eqnarray}\label{ME8}
\hat{H}_{\text{tot}}^\text{M}&\simeq&\omega_{\sigma}\hat\sigma^\dag\hat\sigma
+\omega_{r}\hat{a}^\dag \hat a+\lambda(\hat a\hat\sigma^\dag+\hat{a}^\dag\hat\sigma)\notag\\
&+&\Omega(\hat\sigma e^{i\omega_{L}t}+\hat\sigma^\dag e^{-i\omega_{L}t}).
\end{eqnarray}
This is the well-known driven Jaynes-Cummings (JC) Hamiltonian.

Next, based on the driven JC Hamiltonian, we examine the multi-phonon interactions
in the strong driving limit ($\Omega\gg\lambda$).
Under the resonance driving condition ($\omega_{\sigma}=\omega_{L}$),
the Hamiltonian in Eq.~(\ref{ME8}) can be further simplified.
In the rotating frame with respect to the driving frequency $\omega_L$,  the Hamiltonian Eq.~(\ref{ME8}) reads
\begin{eqnarray}\label{ME9}
\hat{H}_{\text{tot}}^\text{M}=\Delta_{a}\hat{a}^\dag\hat{a}
+\lambda(\hat{a}\hat{\sigma}^\dag+\hat{a}^\dag\hat{\sigma})
+\Omega(\hat{\sigma}+\hat{\sigma}^\dag),
\end{eqnarray}
with $\Delta_{a}=\omega_{r}-\omega_{L}$.

\begin{figure}
\includegraphics[scale=0.24]{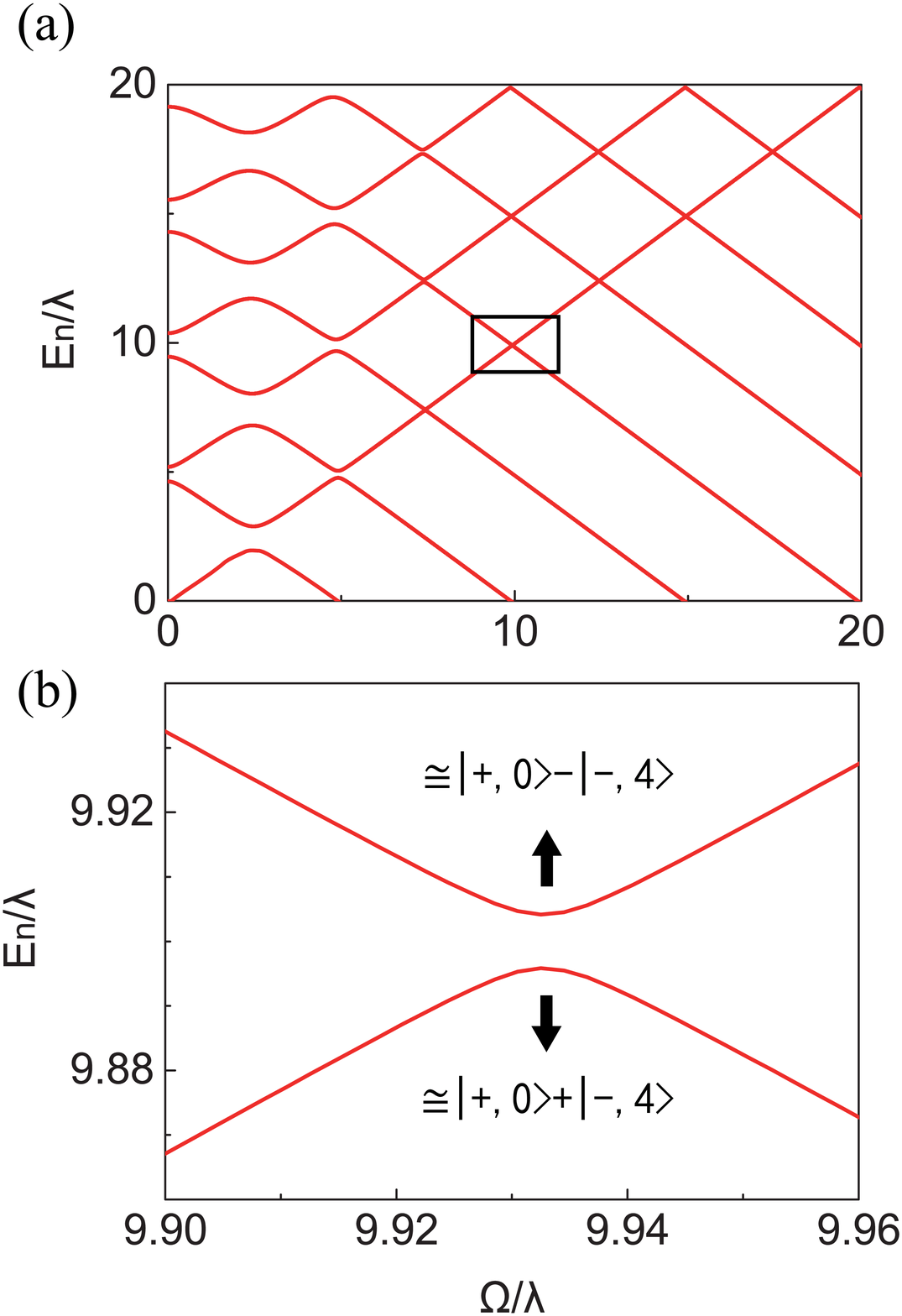}
\caption{\label{fig2a}(Color online) (a) The curve of the energy eigenvalues
with the pumping $\Omega$ increasing from $0$ to $20\lambda$.
The result is calculated from Eq.~(\ref{ME9}) at $\Delta_{a}=5\lambda$.
This plot only contains the second to the ninth energy eigenvalues.
(b) Enlarged view of the spectral region delimited by a square in (a). }
\end{figure}

To show how the multi-phonon transition appears,
we fix the value of $\Delta_{a}$ ($\gg\lambda$) and increase the pumping.
Figure~3(a) shows the energy eigenvalues as a function of $\Omega$ at $\Delta_{a}=5\lambda$.
The obvious avoided crossing arises around $2\Omega=n\Delta_{a}$,
where the resonant transition between the
ground-state and the $n$-phonon state occurs.
In  Fig.~3(b), we show the resonant transition between the states $|+,0\rangle$ and $|-,4\rangle$.
This resonance condition can be well understood in the dressed state basis,
where the effective TLS is dressed with the dressed states $|\pm\rangle=(|d\rangle\pm|b\rangle)/\sqrt{2}$.
The Hamiltonian now reads
\begin{eqnarray}\label{ME10}
\hat{H}_{\text{tot}}^\text{M}&=&\Delta_{a}\hat{a}^\dag\hat{a}+\Omega\hat{\tilde{\sigma}}_{z}\notag\\
&+&\frac{\lambda}{2}
[\hat{a}^\dag(\hat{\tilde{\sigma}}-\hat{\tilde{\sigma}}^\dag+\hat{\tilde{\sigma}}_{z})+\text{H.c.}],
\end{eqnarray}
with $\hat{\tilde{\sigma}}=|-\rangle\langle+|,\hat{\tilde{\sigma}}^\dag,\hat{\tilde{\sigma}}_{z}$
the new Pauli operators.
The strong driving  condition allows us to take
the interaction term (the third term in the right) as a perturbation $\hat{V}$,
and through adjusting the tunable parameters, the high order resonant interaction can stand out
and dominate the coupled system.
Thus, for $n$-phonon resonance,
we can obtain a formalized effective Hamiltonian as
\begin{eqnarray}\label{ME11}
\hat{H}_{\text{eff}}^\text{M}=\Delta_{a}\hat{a}^\dag\hat{a} +\Omega\hat{\tilde{\sigma}}_{z}
+\lambda^{(n)}(\hat{a}^n\hat{\tilde{\sigma}}^\dag+\hat{a}^{\dag n}\hat{\tilde{\sigma}}),
\end{eqnarray}
where $\lambda^{(n)}$ is the corresponding effective $n$-phonon coupling rate.

The $n$-phonon coupling rate has an analytical expression. To derive it,
we employ the Rayleigh-Schrodinger perturbation theory
\cite{hubavc2010brillouin}.
We consider the initial state $|+,0\rangle$ with energy $E_{i}$ and the final state $|-,n\rangle$,
by defining the projection operator $\hat{P}=|+,0\rangle\langle +,0|+|-,n\rangle\langle -,n|$
and the propagator (resolvent) $\hat{K}=\sum\frac{|q\rangle\langle q|}{E_{i}-E_{q}}$
(the summation runs over all possible intermediate states).
The correction to the initial and final states is expressed by
\begin{eqnarray}\label{ME12}
\hat{V}_{\text{corr}}=\hat{P}(\hat{V}+\hat{V}\hat{K}\hat{V}
+\hat{V}\hat{K}\hat{V}\hat{K}\hat{V}+\dots)\hat{P},
\end{eqnarray}
with the diagonal matrix elements representing the shifts
and the nondiagonal matrix elements representing the transition rates.
For the energy shifts, the leading order is $\lambda^2/4\Omega$,
which is about two orders of magnitude smaller than the free energy ($\Omega\sim5\lambda$).
Thus, the resonance condition $2\Omega=n\Delta_{a}$ is approximately satisfied.

Then the $n$-phonon coupling rate is given by
\begin{eqnarray}\label{ME13}
\lambda^{(n)}=\frac{\langle+,0|\hat{V}_{\text{corr}}|-,n\rangle}{(\sqrt{n})!}.
\end{eqnarray}
Only the leading order is kept and the $n$-phonon coupling rate is calculated in the subspace spanned by the states
$\{|+,0\rangle,|-,n\rangle,|+,1\rangle,|-,1\rangle,\cdots,|+,n-1\rangle,|-,n-1\rangle\}$.
We note that each intermediate state provides a negative sign to the $n$-phonon coupling rate.
For intermediate states having $m$ phonons, $\frac{1}{|E_i-E_q|}$ are $\frac{n}{2m\Omega}$ and $\frac{n}{2(n-m)\Omega}$
for states $|m,+\rangle$ and $|m,-\rangle$, respectively.
Then we obtain the analytical expression
\begin{eqnarray}\label{ME14}
\lambda^{(n)}&\approx&(-1)^{n-1}(\frac{\lambda}{2})^n\prod_{1\leq m\leq n-1}[\frac{n}{2m\Omega}+\frac{n}{2(n-m)\Omega}]\notag\\
&=&\frac{(-1)^{n-1}\lambda^n}{2(n-1)!^2}(\frac{n^2}{4\Omega})^{n-1}.
\end{eqnarray}

In general, the coupling strength of the second sideband can reach $2\pi\times10$ kHz, and the third sideband can reach $2\pi\times 1$ kHz.
They are 3 orders of magnitude higher than what can
be achieved with the direct second- or third-order coupling between a single spin and the mechanical mode
\cite{PhysRevLett.121.123604}.

\subsection{The Lamb-Dicke sideband}

Another realizable scheme to control the $n$-phonon transition is pioneered in trapped ions.
\cite{RevModPhys.75.281}.
The interaction between single trapped ions and light field in the Lamb-Dicke regime
gives rise to the resolved sideband, including carrier, red and blue sidebands.
The same idea can be applied to solid-state systems
\cite{PhysRevX.6.041060,PhysRevLett.116.143602,PhysRevX.8.041027,PhysRevB.82.165320,PhysRevB.88.064105}.
In this section, we investigate the multi-phonon sideband transition in a spin-mechanical system.

We consider the case where a single driving field is applied
with $\vec{B}_\text{dr}=B_{x}\cos{\omega_{d}t}\vec{e}_{x}$,
and a stable magnetic field is applied to induce splitting
between the states $\vert -1\rangle$ and $\vert +1\rangle$.
The total Hamiltonian in the Lamb-Dicke regime is written as
\begin{eqnarray}\label{ME15}
\hat{H}_{\text{tot}}^\text{LD}&=&D{\hat{S}}_{z}^{2}+\omega_{r}\hat{a}^\dag\hat{a}
+\frac{\delta_{B}}{2}\hat{S}_{z}\notag\\
&+&\lambda(\hat{a}+\hat{a}^\dag)\hat{S}_{z}
+2\sqrt{2}\Omega\cos{\omega_{d}t}\hat{S}_{x},
\end{eqnarray}
where $\delta_{B}=2g_{s}\mu_{b}(B_{0}+B_{\text{tip}}(h))$ is the static splitting,
and $2\sqrt{2}\Omega=g_{s}\mu_{b}B_{x}$ is the amplitude of the external driving field.

In this case, we choose $\{|0\rangle,|-1\rangle\}$ as the effective qubit without loss of generality, as shown in Fig.~2(c).
Under the rotating wave approximation, the total Hamiltonian is reduced to
\begin{eqnarray}\label{ME16}
\hat{H}_{\text{tot}}^\text{LD}&\simeq&\omega_{\sigma}\hat\sigma^\dag\hat\sigma
+\omega_{r}\hat{a}^\dag \hat a+\lambda(\hat a+\hat{a}^\dag)\hat\sigma_{\text{z}}\notag\\
&+&\Omega(\hat\sigma e^{i\omega_{d}t}+\hat\sigma^\dag e^{-i\omega_{d}t}),
\end{eqnarray}
where $\hat\sigma=|-1\rangle\langle0|$, $\hat\sigma^\dag$ and $\hat\sigma_{\text{z}}$ are the Pauli operators.
In the rotating frame with the driving frequency $\omega_{d}$,
the time independent Hamiltonian is given by
\begin{eqnarray}\label{ME17}
\hat{H}_{\text{tot}}^\text{LD}\simeq\frac{-\delta}{2}\hat\sigma_{\text{z}}
+\omega_{r}\hat{a}^\dag \hat a+\lambda(\hat a+\hat{a}^\dag)\hat\sigma_{\text{z}}
+\Omega(\hat\sigma+\hat\sigma^\dag),
\end{eqnarray}
with $\delta$ the detuning between the effective qubit and the driving field.

Similar to the Mollow regime, we can use the perturbation theory,
and treat the third and the fourth terms in the above equation as the perturbation $\hat{V}$, when $\{\lambda,\Omega\}\ll\{\delta,\omega_{r}\}$.
For example, the two-phonon mediated sideband resonance is the result of the third-order perturbation,
and can be easily expressed as (blue sideband, $\delta\approx 2\omega_{r}$)
\begin{eqnarray}\label{ME18}
\hat{H}_{\text{eff,2}}^\text{LD}=\frac{-\delta}{2}\hat\sigma_{\text{z}}
+\omega_{r}\hat{a}^\dag \hat a+\frac{\Omega}{2!}(\frac{2\lambda}
{\omega_{r}})^2 (\hat{a}^{\dag 2}\hat{\sigma}^\dag+\hat{a}^2\hat{\sigma}).
\end{eqnarray}
We find that, compared to the Mollow regime,
the coupling strength in this regime is suppressed by the factor
$(2\lambda/\omega_{r})^2$, and is very weak in the case of a high frequency resonator.
However, the second-order coupling strength can reach $2\pi\times1$ kHz in the low frequency regime.
It should be pointed out that the coherence of the NV center is not protected in this scheme without dynamical decoupling. Thus, the strong coupling regime is
quite difficult to reach, since the dephasing rate for NV centers is orders of magnitude larger that the two-phonon coupling strength. However, as discussed in Sec.~IV.C., we can find that
even without strong coupling, the Lamb-Dicke scheme can still find some applications that takes advantage of the resonator dissipation and is robust against the spin dephasing.

Another approach to derive the multi-phonon interaction Hamiltonian
is to build the connection between Eq.~[\ref{ME17}] and the Lamb-Dicke model.
By applying the Schrieffer-Wolff transformation
$U=\exp[\frac{\lambda}{\omega_{r}}(\hat{a}^\dag-\hat{a})\hat{\sigma}_{\text{z}}]$,  Eq. (\ref{ME17}) is transformed to
\begin{eqnarray}\label{ME19}
\hat{H}_{\text{tot}}^\text{LD}=\frac{-\delta}{2}\hat\sigma_{\text{z}}
+\omega_{r}\hat{a}^\dag \hat a
+\Omega[e^{\frac{2\lambda}{\omega_{r}}(\hat{a}^\dag-\hat{a})}\hat{\sigma}^\dag+\text{H.c.}].
\end{eqnarray}

The off-resonant sideband transition arises from the phonon-assisted resonant process
in the limit of weak driving, where $\eta=2\lambda/\omega_{r}$ can be viewed as
an effective Lamb-Dicke parameter for this solid-state system.
The red (blue) sidebands appear when $\delta\approx -n\omega_{r}$
($\delta\approx n\omega_{r}$), where the spin is pumped to the upper state,
and the resonator emits (absorbs) n phonons and vice versa.
The effective Hamiltonian of the $n$-phonon interaction is described by (blue sideband)
\begin{eqnarray}\label{ME20}
\hat{H}_{\text{eff}}^{\text{LD}}=\frac{-\delta}{2}\hat\sigma_{\text{z}}
+\omega_{r}\hat{a}^\dag \hat a+\frac{\Omega}{n!}(\frac{2\lambda}
{\omega_{r}})^n (\hat{a}^{\dag n}\hat{\sigma}^\dag+\hat{a}^n\hat{\sigma}).
\end{eqnarray}
This approach gives rise to the same result as that derived from the perturbation theory.
Note that the Lamb-Dicke model can also be realized in the spin-mechanical system based on strain coupling
\cite{PhysRevX.6.041060,PhysRevLett.116.143602,PhysRevX.8.041027,PhysRevB.82.165320}.

\section{Applications}

The multi-phonon interaction featuring nonlinearity is widely studied for
the preparation of nonclassical motion states
\cite{PhysRevLett.121.123604,PhysRevA.88.023817,Munoz2014Emitters,munoz2018filtering},
which may have potential applications in quantum information,
quantum metrology and quantum biology
\cite{Giovannetti1330}.
In this section, we discuss some applications
based on the multi-phonon interaction.

\subsection{Jumping Schrodinger cat states}

\begin{figure}
\includegraphics[scale=0.24]{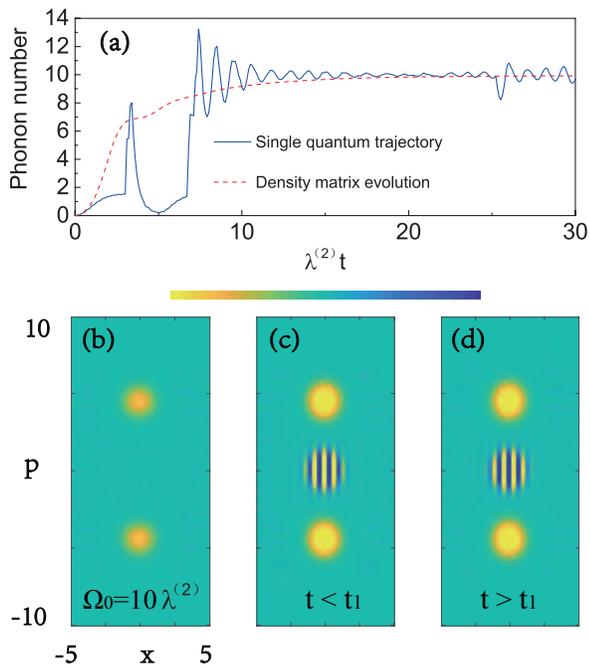}
\caption{\label{fig2a}(Color online) (a) Time evolution  of the phonon number.
(b) Wigner function of the mechanical mode at the steady state given by the density matrix.
(c) and (d) Wigner function of the mechanical mode given by the wave function of a single trajectory.
This two pictures depict a jumping cat at times
before and after a phonon damping process at time $\text{t}_{1}$.
Here, $\lambda^{(2)}=2\times10^{-2}\lambda$, $\Omega_{0}=10\lambda^{(2)}$, $(n_{\text{th}}+1)\gamma_{m}=0.5\lambda^{(2)}$
and $\gamma_{s}=0.05\lambda^{(2)}$.
The numerical results are based on Eq.~(\ref{ME21}) and Eq.~(\ref{ME22}).}
\end{figure}

The Schrodinger cat state is a macroscopic quantum superposition state close to the classical limit.
The two-phonon process is useful for preparing
Schrodinger cat states, as shown in several previous works
\cite{PhysRevLett.121.123604,PhysRevA.88.023817,PhysRevA.49.2785}.
Here, we show that the competition between the two-phonon process and pumping
together with damping makes the system evolve into the mixture of two Schrodinger cat states.
Moreover,  the transient dynamics performs a sustained “jumping cat” with the quantum jump approach.

As for two-phonon process acquired by the second Mollow sideband at $2\Omega=2\Delta_{a}$,
we can acquire the pumping by adding a driving field
along the $z$ direction, with the frequency  equal to the mechanical frequency
and the amplitude equal to the spin-mechanical coupling strength.
Then the effective Hamiltonian in the interaction picture
can be written as
\begin{eqnarray}\label{ME21}
\hat{H}_{I}=\lambda^{(2)}(\hat{a}^2\hat{\tilde{\sigma}}^\dag+\hat{a}^{\dag 2}\hat{\tilde{\sigma}})
+\Omega_{0}(\hat{\tilde{\sigma}}^\dag+\hat{\tilde{\sigma}}),
\end{eqnarray}
where $\Omega_{0}$ is the effective driving strength in the second sideband.

In order to describe the dynamics of the system under dissipation,
we turn to both master equation approach
\cite{Carmichael1999Statistical}
and quantum jump approach
\cite{RevModPhys.70.101}.
The master equation with Lindblad terms is
\begin{eqnarray}\label{ME22}
\dot{\hat{\rho}}\simeq-i[\hat{H}_{I},\hat{\rho}]
+\gamma_{s}D[\hat{\tilde{\sigma}}_{z}]\hat{\rho}+(n_{\text{th}}+1)\gamma_{m}D[\hat{a}]\hat{\rho},
\end{eqnarray}
with $D[\hat{O}]\hat{\rho}=\hat{O}\hat{\rho}\hat{O}^\dag
-\frac{1}{2}\hat{O}^\dag\hat{O}\hat{\rho}-\frac{1}{2}\hat{\rho}\hat{O}^\dag\hat{O}$
for a given operator $\hat{O}$.
We assume that the system is actively cooled to a thermal
phonon population close to zero, which can be done, for instance,
by means of laser cooling or using another spin qubit
\cite{PhysRevLett.92.075507,OConnell2010Quantum,Teufel2011Sideband,Chan2011Laser,
rabl2009strong,PhysRevB.82.165320,PhysRevB.88.064105,PhysRevLett.121.123604}.
Therefore, we can exclude the term $D[\hat{a}^\dag]\hat{\rho}$ from the master equation.
The dephasing rate of the spin is $\gamma_{s}$,
the damping rate of the resonator is $\gamma_{m}$,
and the thermal phonon number is $n_{\text{th}}$ in the absence of cooling,
which is determined by the environmental temperature.
Because the Schrodinger cat state is parity protected, it's less sensitive to dephasing.
Furthermore, with a proper design, the coupling strength $\lambda$ can exceed $2\pi\times100$ kHz,
while the dephasing rate is about one order of magnitude smaller than the second-order coupling strength.
Thus the effect of the dephasing can be neglected.

When the damping is engineered such that it is smaller than the second-order coupling strength,
the system will evolve towards the dark state $|\psi_{d}\rangle$
even if $D[a]|\psi_{d}\rangle\langle\psi_{d}|\neq0$.
The dark state condition is
\begin{eqnarray}\label{ME23}
\hat{H}_{I}|\psi_{d}\rangle=0,
\end{eqnarray}
where the mechanical dark state $|\psi_{a}\rangle$ satisfies the equation
\begin{eqnarray}\label{ME24}
\hat{a}^2|\psi_{a}\rangle=-\Omega_{0}/\lambda^{(2)}|\psi_{a}\rangle.
\end{eqnarray}
It's proved that the solutions are two states with opposite parity known as Schrodinger cat states
\begin{eqnarray}\label{ME25}
|\psi_{a}\rangle_{0}&=&N_{e}^{-1/2}(|\beta\rangle+|-\beta\rangle)\\
|\psi_{a}\rangle_{1}&=&N_{o}^{-1/2}(|\beta\rangle-|-\beta\rangle)\label{ME26},
\end{eqnarray}
with $\beta=\sqrt{-\Omega_{0}/\lambda^{(2)}}$,
$N_{e}^{-1/2}=2[1+\exp(-2|\beta|^2)]$, and $N_{o}^{-1/2}=2[1-\exp(-2|\beta|^2)]$.
The phonon number of the steady state is of $|\beta|^2$.

Figure 4 displays the transient dynamics of the mechanical resonator towards the steady state under the two-phonon
Hamiltonian (\ref{ME21}). Fig.~4(a) shows the phonon population versus time based on the density matrix and a single quantum trajectory.
Here, the mechanical resonator is initially in the ground state and the spin is initially in the upper state.
Fig.~4(b) shows the Wigner function of the mechanical resonator at the
steady state computed from the density matrix approach,  where  just the phase bimodality  appears.
However, when using Monte carlo simulations with individual quantum trajectories
in the photon-counting configuration, we find that the bimodal steady state consists of a cat state that undergoes
a random phase flip, as shown in Fig.~(c) and (d).
The interference fringes are observed in the middle of the bimodality,
which are indicative of quantum coherence and the phase flip
reveals the jump between the two cat-like states.

\subsection{Preparing multi-phonon Fock states}

The phononic Fock state is fundamentally important  in quantum technology involving phonons.
In this section, we show how to prepare the multi-phonon Fock states with this hybrid spin-mechanical system.
Here we choose the scheme that allows the Mollow dynamics.
The dynamics of the system is governed by the following
master equation
\begin{eqnarray}\label{ME27}
\dot{\hat{\rho}}&=&-i[\hat{H}_{\text{eff}}^\text{M},\hat{\rho}]
+\gamma_{s}D[\hat{\tilde{\sigma}}_{z}]\hat{\rho}\notag\\
&+&n_{\text{th}}\gamma_{m}D[\hat{a}^\dag]\hat{\rho}+(n_{\text{th}}+1)\gamma_{m}D[\hat{a}]\hat{\rho}.
\end{eqnarray}
Here, we reserve the absorption term proportional to thermal phonon number.

\begin{figure}
\includegraphics[scale=0.165]{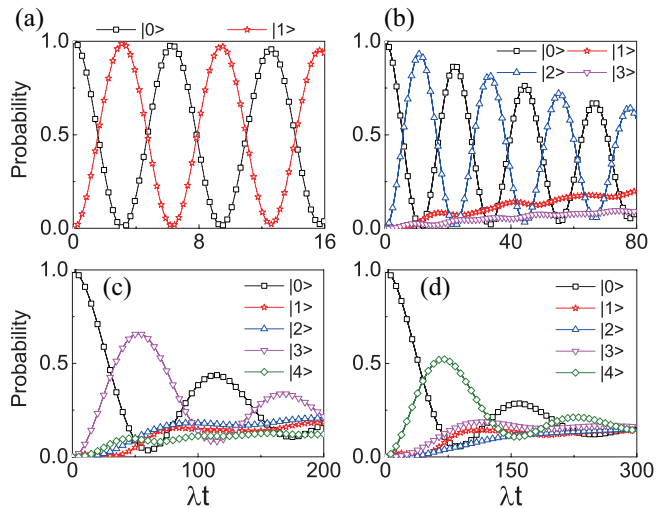}
\caption{\label{fig:wide}(Color online)  Time evolution of the probabilities
for the resonator being in certain number states $\vert n\rangle$, with the initial state $|\psi(t=0)\rangle=|+,0\rangle$.
The relevant dissipation parameters are
$\gamma_{s}=10^{-3}\lambda$, $\gamma_{m}=5\times 10^{-5}\lambda$, and $n_{th}=40$.
(a) One-phonon resonance, with $\Delta_{a}=10\lambda$ and $\Omega=5\lambda$;
(b) Two-phonon resonance, with $\Delta_{a}=5\lambda$ and $\Omega=5\lambda$;
(c) Three-phonon resonance, with $\Delta_{a}=5\lambda$ and $\Omega=7.5\lambda$;
(d) Four-phonon resonance, with $\Delta_{a}=3\lambda$ and $\Omega=6\lambda$. }
\end{figure}

Figure~5 shows the  dynamics of the system on $n$-phonon resonance, with $n=1,2,3,4$, respectively.
The numerical results are calculated  based on Eq.~(\ref{ME11}) and Eq.~(\ref{ME27}).
Here the system is initialized in the state $|0,+\rangle$.
The coupling strength is given by Eq.~(\ref{ME14}).
Fig.~5(a) plots the probability of the resonator being in the states
$|0\rangle$ and $|1\rangle$ on one-phonon resonance,
with the coupling rate $\lambda^{(1)}=0.5\lambda$.
We find that the probability for the one-phonon state can be close to unity
over several Rabi oscillations.
Fig.~5(b) plots the probability for the resonator being in the states
$|n\rangle$ on two-phonon resonance, with $n$ from $0$ to $3$.
The coupling rate is  $\lambda^{(2)}=-0.1\lambda$.
We find that, the probability of having two phonons is sizable, while the probability of having
one phonon is more than one order of magnitude smaller. Therefore, this relatively
high probability of the two-phonon state indicates that the two-phonon interaction dominates the dynamics.

Fig.~5(c) and (d) show the Rabi flopping corresponding to the Mollow regime, where
the three- and four-phonon interactions occur. For the three-phonon resonance shown in Fig.~5(c), the coupling rate is $\lambda^{(3)}=0.01125\lambda$, while for the four-phonon resonance shown in Fig.~5(d), the coupling rate is $\lambda^{(4)}\approx-0.004\lambda$. From Fig.~5(c) and (d), we find that the probability of three-phonon state can reach $0.7$, while  the probability of four-phonon state can still exceed $0.5$ under
realistic parameters.

\subsection{$n$-phonon correlations}

The $n$-photon resonance transition can be used to realize emitters
that releases their energy in groups (or ‘bundles’) of $n$ photons in cavity quantum electrodynamics schemes
\cite{Munoz2014Emitters,munoz2018filtering}.
On the other hand, some work suggests the phonon laser
when higher order phonon processes become relevant
\cite{PhysRevB.88.064105}.
Here, we show the quantum correlation of the $n$-phonon with the engineered dissipation,
which simultaneously demonstrates quantum and nonlinear properties.

We provide a controlled effective decay rate for the spin
by an excited-state-mediated process.
In this process, the spin states $|\pm1\rangle$ are optically pumped to the higher electronic level of the color center,
and then they decay back to the states $|\pm1\rangle$ with a rate $\Gamma_{1}$
and to the state $|0\rangle$ with a rate $\Gamma_{0}$ depending on the pumping strength
\cite{rabl2009strong,Rabl2010A}.
Using resonant excitations of appropriately chosen transitions at low temperatures,
the decay rate to the states $|\pm1\rangle$ can approach zero
\cite{Tamarat2008}.
Thus we equivalently obtain the required decay from the spin state $|\pm1\rangle$ to  the state $|0\rangle$.
Here, we consider the Lamb-Dicke scheme when the blue sideband condition is satisfied.
Note that the discussion is also applied to the Mollow regime.

The system under an engineered decay is described by the following master equation
\begin{eqnarray}\label{ME28}
\dot{\hat{\rho}}&\simeq&-i[\hat{H}_{\text{eff}}^\text{LD},\hat{\rho}]+(n_{\text{th}}+1)\gamma_{m}D[\hat{a}]\hat{\rho}\notag\\
&+&\gamma_{s}D[\hat{\sigma}_{z}]\hat{\rho}+\Gamma_{0}D[\hat{\sigma}]\hat{\rho}.
\end{eqnarray}
The incoherence term $D[\hat{a}^\dag]\hat{\rho}$ is excluded
when the system is actively cooled.
The typical decoherence rate of NV center ($\sim2\pi\times100$ kHz) is used here.
We note that the irrelevant single-phonon relaxing process given by
the off-resonant population is suppressed.

To characterize these quantum correlations,
we employ the generalized second-order correlation function $g_{n}^{(2)}(\tau)$ as \cite{PhysRev.130.2529,Munoz2014Emitters}
\begin{eqnarray}\label{ME29}
g_{n}^{(2)}(\tau)=\frac{\langle\hat{a}^{\dag n}(0)\hat{a}^{\dag n}(\tau)
\hat{a}^{n}(\tau)\hat{a}^{n}(0)\rangle}{\langle(\hat{a}^{\dag n}\hat{a}^{n})(0)\rangle
\langle(\hat{a}^{\dag n}\hat{a}^{n})(\tau)\rangle}.
\end{eqnarray}
As a result,
$g_{n}^{(2)}(\tau)\leq g_{n}^{(2)}(0)$ ($g_{n}^{(2)}(\tau)> g_{n}^{(2)}(0)$)
indicates the bunching (antibunching) of the phonon field, while $g_{n}^{(2)}(0)> 1$ ($g_{n}^{(2)}(0)< 1$) indicates the super-Poisson (sub-Poisson) distribution.

\begin{figure}
\includegraphics[scale=0.19]{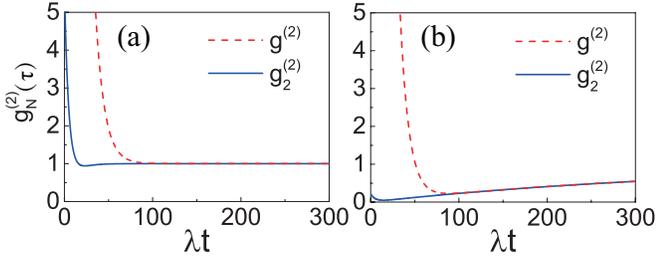}
\caption{\label{fig:wide}(Color online)  Second-order phonon correlations
at the $N=1$ (dashed, red) and $N=2$ (solid, blue) level at two-phonon resonance,
with quadratic coupling strength
$\lambda^{(2)}=10^{-2}\lambda$, dephasing rate $\gamma_s=10^2\lambda^{(2)}$ and damping rate $(n_{\text{th}}+1)\gamma_{m}=10\lambda^{(2)}$.
(a) Short effective lifetime of the spin qubit with $\Gamma_{0}=10\lambda^{(2)}$;
(b) Long effective lifetime of the spin qubit with $\Gamma_{0}=0.25\lambda^{(2)}$. }
\end{figure}

Figure 6 shows the phonon correlation at two-phonon resonance
in terms of standard and generalized second-order correlation function
both in the limit of short and long effective lifetime of the spin qubit. The numerical results are
based on Eq.~(\ref{ME18}) and Eq.~(\ref{ME28}). For short lifetime,
the phonon field is bunched and has a super-Poisson distribution of single phonons and phonon pairs.
This limit may allow the creation of phonon laser with $n$-phonon released.
On the other hand, in the long lifetime limit,
we observe the phenomena  of antibunching and sub-Poisson distribution of phonon pairs,
which fully uncovers the quantum and nonlinear features
though the second-order correlations.

To give a better understanding of the two opposite distributions of the photon field,
we briefly analyze the dynamics from the viewpoint of  the  quantum jump approach.
Without loss of generality, we consider the condition at two-phonon resonance.
The system is first prepared in the state $|-,0\rangle$,
where the spin qubit is in the lower state and the mechanical mode is in the ground state.
With the quadratic interaction, the system evolves towards the state $|+,2\rangle$.
However, due to the large damping rate,
the system collapses and relaxes (emits) a cascade of two-phonon in a short period of time,
and now stays in the state $|+,0\rangle$, where the quadratic interaction doesn't work.
Until the spin decays into the lower state,
the system is brought back to its starting point and
then repeats the dissipation process, relaxing the cascade of two-phonon.
The effective lifetime of the spin determines the non-interaction window,
which dominates the time interval between two neighbour cycles.
Thus phonon pairs exhibit antibunching for long-lived qubit,
while they are super-Poisson distributed for short-lived qubit.

\section{The feasibility of this scheme}

We now consider the experimental feasibility of this proposal and discuss the relevant parameters in our scheme.
We choose the silicon cantilever as the mechanical resonator with dimensions (l,w,t)=(3.47,0.05,0.05) $\mu$m
\cite{Arcizet2011A,rabl2009strong,Rabl2010A},
the Young's modulus $E\sim 1.3\times10^{11}$ Pa,
and the mass density $\varrho\sim 2.33\times10^{3}$ $kg/m^{3}$.
Then, by using the expression $\omega_{r}\sim3.516\times(t/l^{2})\sqrt{E/12\varrho}$
\cite{RevModPhys.67.249},
the fundamental frequency is $\omega_{r}\sim2\pi\times5$ MHz.
With the resonator's effective mass $m_{\text{eff}}=\varrho lwt/4$,
the zero-field fluctuation $a_{0}$ is around $5.7\times10^{-13}$ m.
Also, a sharp magnetic tip is used to produce a magnetic gradient of $G_{m}\sim10^{7}$ T/m
at a distance $h=25$ nm away from the tip
\cite{Mamin2007Nuclear}.
Thus, the magnetic coupling strength between
the resonator and the NV center is about $\lambda\sim2\pi\times150$ kHz
\cite{rabl2009strong}.
Because larger cantilevers are more realistic in experiment,
we specially consider a single-crystal diamond nanomechanical resonator with dimensions (l,w,t)=(20,8,0.8) $\mu$m
\cite{Tao2014Single}.
The corresponding resonance frequency and the zero-field fluctuation
are about $2\pi\times 5.7$ MHz and $3.6\times10^{-15}$ m, respectively.
The coupling strength $\lambda$ is then reduced to the order of kHz, and
the multi-phonon interaction is hard to  reach the strong coupling regime.
So a small size cantilever is more desirable in our model.

The experiment is carried out in dilution refrigerator
with the temperature $T\sim10$ mK.
The thermal phonon number is about $n_{\text{th}}=1/(e^{\hbar\omega_{r}/k_{B}T}-1)\sim40$.
On the other hand, the mechanical resonator can be fabricated with high quality factor $Q>10^6$
\cite{Ovartchaiyapong2012High,Tao2014Single}.
The coherence time of the mechanical resonator used in the numerical calculation is smaller than $5$ ms.
As for single NV centers, the decoherence results from
the spin-lattice relaxation and dephasing.
The relaxation time $T_{1}$ can be several
minutes at low temperatures
\cite{PhysRevLett.101.047601}.
Moreover, the dressed spin states are insensitive to perturbations from the nuclear-spin bath
\cite{rabl2009strong,PhysRevB.92.224419},
and owning to the development of dynamical decoupling techniques,
the dephasing time $T_{2}$ can exceed $2$ ms in experiment
\cite{Balasubramanian2009Ultralong,Hason2008Coherent,Du2009Preserving}.
Thus, we choose $T_2\sim10$ ms for the Mollow scheme and $T_2\sim10$ $\mu$s for the Lamb-Dicke scheme.

\begin{figure}
\includegraphics[scale=0.24]{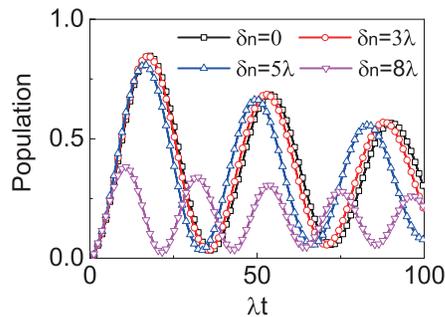}
\caption{\label{fig:wide}(Color online)
The population of the two-phonon state on two-phonon resonance under different noise strength.
The detuning is $\Delta\approx2\pi\times250$ MHz and the driving amplitude is $\Omega_x\approx2\pi\times25$ MHz.
with $\lambda\approx2\pi\times100$ kHz, $\Delta_a=8\lambda$, $\Omega=7.94\lambda$, $\gamma_{s}=10^{-3}\lambda$,
$\gamma_{m}=5\times 10^{-5}\lambda$ and $n_\text{th}=40$.
}
\end{figure}

We now study the feasibility of the schemes under some noise.
We consider the static drift in the magnetic field and take the Mollow regime as an example.
The spin Hamiltonian is modified under that drift,
$\hat{H}_\text{spin}'=\hat{H}_\text{spin}+\delta_n\hat{S}_{z}$, with $\delta_n$ the random amplitude.
The eigenfrequencies in the dressed-state picture are subsequently shifted and it
may prevent resonant interactions between the spin and the resonator mode.
To see how the results change with respect to the noise strength,
we consider the case of the two-phonon resonance transition.
In Fig.~7 we plot the two-phonon state population as a function of time
on two-phonon resonance ($\Delta_a\approx\Omega$) under different noise strength,
with the detuning $\Delta\approx2\pi\times250$ MHz and the driving amplitude $\Omega_x\approx2\pi\times25$ MHz.
The numerical results are based on Eq.~(\ref{ME8}) and Eq.~(\ref{ME27}),
with  a free energy term $(\delta\omega_{bd}/2)\hat{\sigma}_z$ added to the Hamiltonian (\ref{ME8}).
We find that the perturbations induced by the static drift in the magnetic field are negligible for $\delta_n\leq5\lambda$,
but the result significantly changes with respect to noise strength $\delta_n=8\lambda$.
Moreover, the perturbations can be further suppressed by increasing the detuning $\Delta$.
So the proposed schemes are feasible if the external noise is not too large.

\section{Conclusion}

In summary, we have studied the multi-phonon interactions between
a single defect center in diamond and a single phononic mode.
The realization of such interaction is discussed in detail by deriving an effective Hamiltonian to describe it.
We show that, when the spin qubit  is driven,
the coupling strength in  the second sideband can reach $2\pi\times10$ kHz
while the coupling strength in the third sideband can reach $2\pi\times1$ kHz,
allowing the coherent manipulation of $n$ phonons in the spin-mechanical system.
Based on this multi-phonon interaction, we discuss the preparation of quantum states
and explore the quantum correlations of the mechanical motion.
This general work utilizes defect centers to control the phononic mode,
and can be applied to other hybrid system to further explore the quantum acoustics.

\section*{Acknowledgments}
This work is supported by the NSFC under Grant
No. 11774285 and the
Fundamental Research Funds for the Central Universities.

%
\end{document}